\documentclass[graybox]{svmult}

\newcommand{\things}{\textit{'things' }}

\usepackage{mathptmx}       
\usepackage{helvet}         
\usepackage{courier}        
\usepackage{type1cm}        
\usepackage{makeidx}         
\usepackage{graphicx}        
\usepackage{multicol}        
\usepackage[bottom]{footmisc}
\usepackage{color}
\usepackage{url}
\usepackage{seqsplit}
\usepackage{graphicx}
\usepackage{caption}
\usepackage{subfigure}
\usepackage{pdfpages}

\makeindex             

\begin{document}

\title*{Context-aware Dynamic Discovery and Configuration of \textit{`Things'} in Smart Environments}

\titlerunning{Configuration of \textit{`Things'} in Smart Environments}
\author{Charith Perera, Prem Prakash Jayaraman, Arkady Zaslavsky, Peter Christen, and Dimitrios Georgakopoulos}

\authorrunning{Perera et al.}

\institute{Charith Perera, Prem Jayaraman, Arkady Zaslavsky, Dimitrios Georgakopoulos \at CSIRO Computational Informatics, Canberra, ACT 2601, Australia, \email{firstname.lastname@csiro.au}
\and Charith Perera, Peter Christen \at Research School of Computer Science, The Australian National University, Canberra, ACT 0200, Australia, \email{firstname.lastname@anu.edu.au}}
%
%
\maketitle

\abstract*{ The Internet of Things (IoT) is a dynamic global information  network consisting of Internet-connected objects, such as RFIDs, sensors, actuators, as well as other instruments and smart appliances that are becoming an integral component of the future Internet. Currently, such Internet-connected objects or \textit{`things'} outnumber both people and computers connected to the Internet and their population is expected to grow to 50 billion in the next 5 to 10 years. To be able to develop IoT applications, such \textit{`things'} must become dynamically integrated into emerging information networks supported by architecturally scalable and economically feasible Internet service delivery models, such as cloud computing. Achieving  such integration through discovery and configuration of \textit{`things'} is a challenging task. Towards this end, we propose a Context-Aware Dynamic Discovery of {Things} (CADDOT) model. We have developed a tool \textit{SmartLink}, that is capable of discovering sensors deployed in a particular location despite their heterogeneity. \textit{SmartLink} helps to establish the direct communication between sensor hardware and cloud-based IoT middleware platforms. We address the challenge of heterogeneity using a plug in architecture. Our prototype tool is developed on an Android platform. Further, we employ the Global Sensor Network (GSN) as the IoT middleware for the proof of concept validation. The significance of the proposed solution is validated using a test-bed that comprises 52 Arduino-based Libelium sensors.}

\abstract{ The Internet of Things (IoT) is a dynamic global information  network consisting of Internet-connected objects, such as RFIDs, sensors, actuators, as well as other instruments and smart appliances that are becoming an integral component of the future Internet. Currently, such Internet-connected objects or \textit{`things'} outnumber both people and computers connected to the Internet and their population is expected to grow to 50 billion in the next 5 to 10 years. To be able to develop IoT applications, such \textit{`things'} must become dynamically integrated into emerging information networks supported by architecturally scalable and economically feasible Internet service delivery models, such as cloud computing. Achieving  such integration through discovery and configuration of \textit{`things'} is a challenging task. Towards this end, we propose a Context-Aware Dynamic Discovery of {Things} (CADDOT) model. We have developed a tool \textit{SmartLink}, that is capable of discovering sensors deployed in a particular location despite their heterogeneity. \textit{SmartLink} helps to establish the direct communication between sensor hardware and cloud-based IoT middleware platforms. We address the challenge of heterogeneity using a plug in architecture. Our prototype tool is developed on an Android platform. Further, we employ the Global Sensor Network (GSN) as the IoT middleware for the proof of concept validation. The significance of the proposed solution is validated using a test-bed that comprises 52 Arduino-based Libelium sensors.
}

\section{Introduction}
\label{sec:Introduction}

The Internet of Things (IoT) \cite{P003} rfirst received attention in the late 20th century. The term was firstly coined by Kevin Ashton \cite{P065} in 1999.  \textit{``The Internet of Things allows people and things\footnote{We use both terms, `\textit{objects}' and `\textit{things}' interchangeably to give the same meaning as they are frequently used in IoT related documentation. Some other terms used by the research community are `smart objects', `devices', `nodes'. Each `thing' may have one or more sensors attached to it.} to be connected Anytime, Anyplace, with Anything and Anyone, ideally using Any path/ network and Any service''} \cite{P019}. As highlighted in the above definition, connectivity among devices is a critical functionality that is required to fulfil the vision of IoT. The following statistics highlight the magnitude of the challenge we need to address. Due to the increasing popularity of mobile devices over the past decade, it is estimated that there are about 1.5 billion Internet-enabled PCs and over 1 billion Internet-enabled mobile devices today. The number of \things connected to the Internet exceeded the number of people on earth in 2008 \cite{P574}. By 2020, there will be 50 to 100 billion devices connected to the Internet \cite{P029}. Similarly, according to BCC Research, the global market for sensors was around \$56.3 billion in 2010. In 2011, it was around \$62.8 billion, and it is expected to increase to \$91.5 billion by 2016, at a compound annual growth rate  (CAGR) of 7.8\%  \cite{P255}.

The above statistics allow us to conclude that the growth rate of sensors being deployed around us is increasing over time and will keep its pace over the coming decade. Over the last few years, we have witnessed many IoT solutions making their way into the market  \cite{P596}. The IoT market has already been fragmented, with many parties competing with a variety of different solutions. Broadly, these IoT solutions can be divided into two segments: sensor hardware-based solutions  \cite{P595} and cloud-based software solutions \cite{P579, P227, P377}. Some products specifically address one segment, while others address both. In this chapter, we propose a Context-Aware Dynamic Discovery of Things (CADDOT) model in order to support the integration of  \textit{`things'} into cloud-based IoT solutions via dynamic discovery and configuration by also addressing the challenge of heterogeneity. We reduce the complexity of the \textit{`things' configuration process} and make it more user friendly and easier to use. One major objective is to support non-technical users by allowing them to configure smart environments without technical assistance.

This chapter makes the following contributions. We propose a model, CADDOT, that can be used to configure sensors autonomously without human intervention in highly dynamic smart environments in the Internet of things paradigm. To support this model, we developed a tool called \textit{SmartLink}. \textit{SmartLink} is enriched with context-aware capabilities so it can detect sensors using  different protocols such as TCP, UDP, Bluetooth and ZigBee.  CADDOT is designed to deal with highly dynamic smart environments where sensors are appearing and disappearing at a high frequency. This chapter also presents the results of experimental evaluations performed using 52 sensors measuring different types of phenomenon and using different communication sequences.

We explain how our model can be used to enrich the existing solutions proposed in the research field. The chapter is organized as follows. We present background information and motivation in Section \ref{sec:Background}. In Section \ref{sec:Functional_Requirements}, we discuss the functional requirements of an ideal IoT configuration process. We discuss related work in Section \ref{sec:Related_Work}. The proposed CADDOT model is introduced in Section \ref{sec:Architectural_Design}. The design decisions we made are justified and compared with alternative options in Section \ref{sec:Design_Decisions}. Implementation details and evaluations are presented in Section \ref{sec:Implementation} and Section \ref{sec:Evaluation_of the_Prototype} respectively. The lessons learnt are discussed in Section \ref{sec:Discussion}. Open challenges are presented in Section \ref{sec:Open_Challenges} and we conclude the chapter in Section \ref{sec:Conclusions} with indications for future work.

\section{Background and Motivation}
\label{sec:Background}

This section briefly highlights the background details of the challenge we address in this chapter. Firstly, we explain the challenges in the smart environment from the perspective of dynamic discovery and configuration of \things. Secondly, we discuss the concept of sensing as a service and its impact on the IoT. At the end, we present the importance of the configuration of \things in the big data domain.

\subsection{Smart Environment}
\label{sec:BM:Smart_Environment}

A smart environment can be defined as \textit{``a physical world that is richly and invisibly interwoven with sensors, actuators, displays, and computational  elements, embedded seamlessly in the everyday objects of our lives, and connected through a continuous network"} \cite{P633}. Smart environments may be embedded with a variety of smart devices of different types including tags, sensors and controllers, and have different form factors ranging from nano to micro to macro sized. As also highlighted by Cook and Das \cite{P634}, device communication using middleware and wireless communication is a significant part of forming a connected environment. Forming smart environments needs several activities to be performed, such as discovery (i.e. exploring and finding devices at a given location), identification (i.e. retrieving information about devices and recognizing them), connection establishment (i.e. initiating communication using a protocol that the device can understand), and configuration. Further, users may combine sensors and services to configure smart environments where actuators are automatically triggered based on conditions \cite{E4}. In smart home environments, Radio Frequency for Consumer Electronics (RF4CE) has been used to perform atuomated configuration of consumer devices \cite{E6}. However, such techniques cannot be used to configure low-level smart \textit{`things'}.

\vspace{-8pt}

\subsection{Sensing as a service}
\label{sec:BM:Sensing_as_a_service}

The sensing-as-a-service  model \cite{ZMP008} provides sensing capabilities as a service similar to other models such as infrastructure-as-a-service (IaaS), platform-as-a-service (PaaS), and software-as-a-service (SaaS). Mobile devices are widely used to collect data from inbuilt or external sensors \cite{E2}.

It envisions that sensor descriptions and capabilities are posted on the Internet so the interested  consumer can get access to the corresponding sensors by paying a fee \cite{ZMP008}. The sensing as a service model is expected to drive the IoT from the business point of view by creating a whole new set of opportunities and values. It has been predicted that individuals as well as, private and public organizations will deploy sensors to achieve their primary objectives \cite{HomeOS, ZMP008}. Additionally, they will share their sensors with others so a collectively value-added solution can be built around them. Such sensor deployments and data collection allows the creation of real-time solutions to address tough challenges in Smart Cities \cite{E5, ZMP008}. In order to support sensor deployments, easy-to-use \things discovery and configuration tools need to be developed. Such a set of tools will stimulate the growth of sensor deployments in the IoT. They will help the non-technical community to become involved in building smart environments efficiently and effectively.

\subsection{Big Data Challenge}
\label{sec:BM:Big_Data}

\begin{figure}[b]
 \centering
 \includegraphics[scale=0.36]{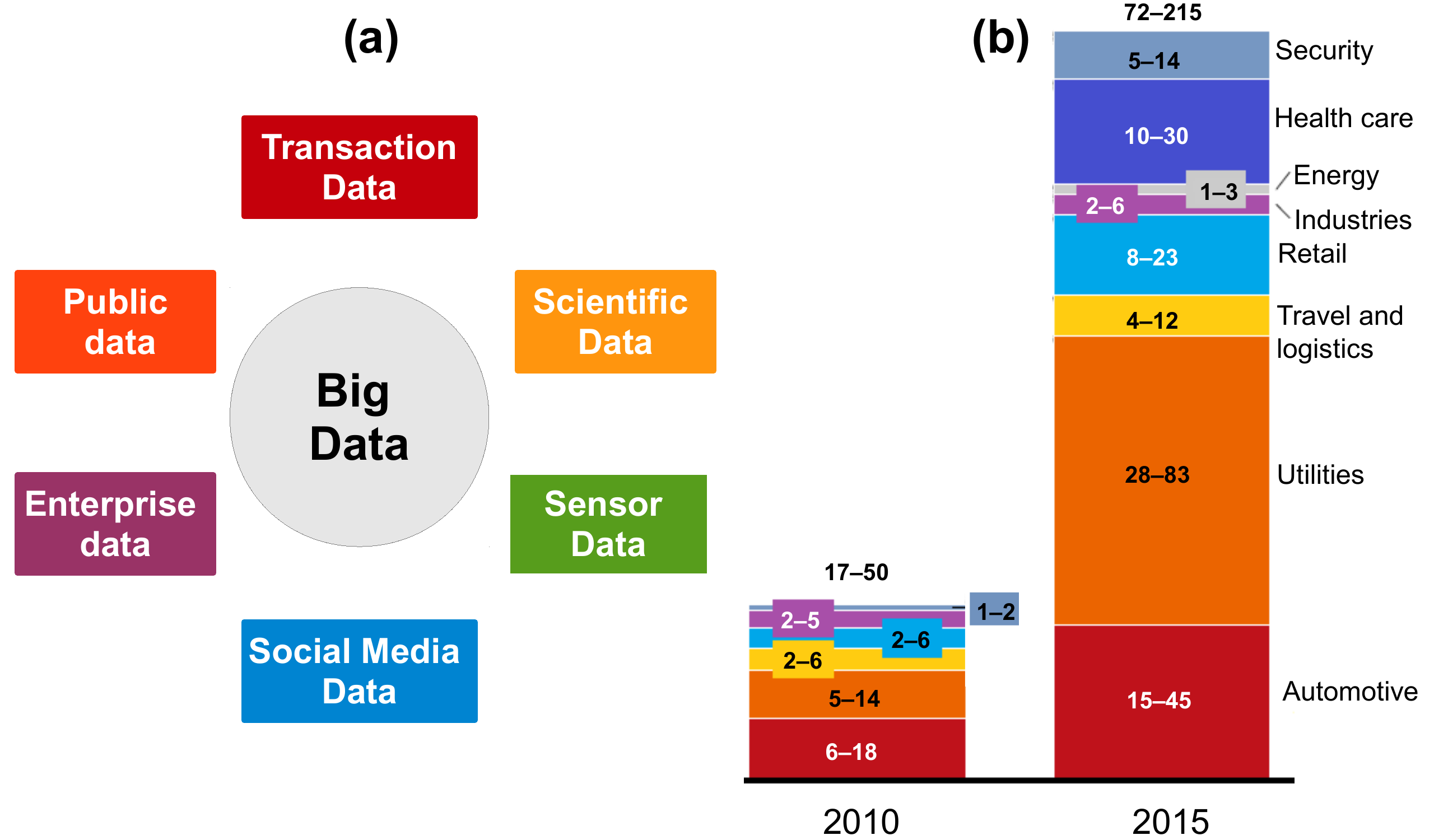}
 \caption{(a) Big Data comprises six categories of data  (b) Data generated from the IoT will grow exponentially as the number of connected nodes increases. Estimated numbers of connected nodes based on different sectors are presented in millions \cite{P504}.}
 \label{Figure:Statistics}
\end{figure}

Big Data \cite{P503} mainly comprises six categories of data, as illustrated in Figure\ref{Figure:Statistics}(a) transaction data, scientific data, sensor data, social media data, enterprise data, and public data. The sensor data category is expected to be generated  by the growing number of sensors deployed in different domains, as illustrated in Figure \ref{Figure:Statistics}. The data streams coming from \things will challenge the traditional approaches to data management and contribute to the emerging paradigm of big data. Collecting  sensor data on a massive scale, which  creates big data, requires easy-to-use sensor discovery  and configuration tools that help to integrate the \things into cloud-based IoT middleware platforms. Big data has been identified as a secondary phase of the IoT, where new sensors are cropping up and organizations are now starting to analyse data, that in some cases, they have been collecting for years.

This work is also motivated by our previous work which focused on utilising mobile phones and similar capacity devices to collect sensor data. In DAM4GSN \cite{ZMP001}, we proposed an application that can be used to collect data from sensors built  into mobile phones. Later, we proposed MoSHub \cite{ZMP005} that allows a variety of different external sensors to be connected to a mobile phone using an extensible plugin architecture. MoSHub also configures the cloud middleware accordingly. Later in MOSDEN \cite{ZMP009}, we developed a complete middleware for resource-constrained mobile devices. MOSDEN is capable of collecting data from both internal and external sensors. It can also apply SQL-based fusing on data streams in real time. As we mentioned earlier, in order to collect data from sensors, first we need to discover and configure the sensors in such a way that the cloud can communicate with them. In our previous efforts, discovery and configuration steps were performed manually. In this chapter, we propose an approach that can be used to discover and configure sensors autonomously.

\vspace{-8pt}
\section{Functional Requirements}
\label{sec:Functional_Requirements}

The \things \textit{configuration process} detects, identifies, and configures sensor hardware and cloud-based IoT platforms in such a way that software platforms can retrieve data from sensors when required. In this section, we identify the importance, major challenges, and factors that need to be considered during a configuration process. The process of sensor configuration in IoT is important for two main reasons. Firstly, it establishes the connectivity between sensor hardware and software systems wich makes it possible to  retrieve data from the  deployed sensor. Secondly, it allows us to optimize the sensing and data communication by considering several factors as discussed below.  Let us discuss the following research problem:\textit{ ‘Why is sensor configuration challenging in the IoT environment?’}. The major factors that make sensor configuration challenging are \textit{1) the number of sensors, 2) heterogeneity, 3) scheduling, sampling rate, communication frequency, 4) data acquisition, 5) dynamicity, and 6) context} \cite{ZMP007}.

\textbf{1) Number of Sensors:} When the number of sensors that need to be configured is limited, we can use manual  or semi-autonomous techniques. However, when the numbers grow rapidly towards millions and billions, as illustrated in Figure \ref{Figure:Statistics}(b), such methods become extremely inefficient, expensive, labour-intensive, and in most situations impossible. Therefore, large numbers have made sensor configuration challenging. An ideal sensor configuration approach should be able to configure sensors autonomously as well as within a very short time period.

\begin{figure}[h]
 \centering
 \includegraphics[scale=0.45]{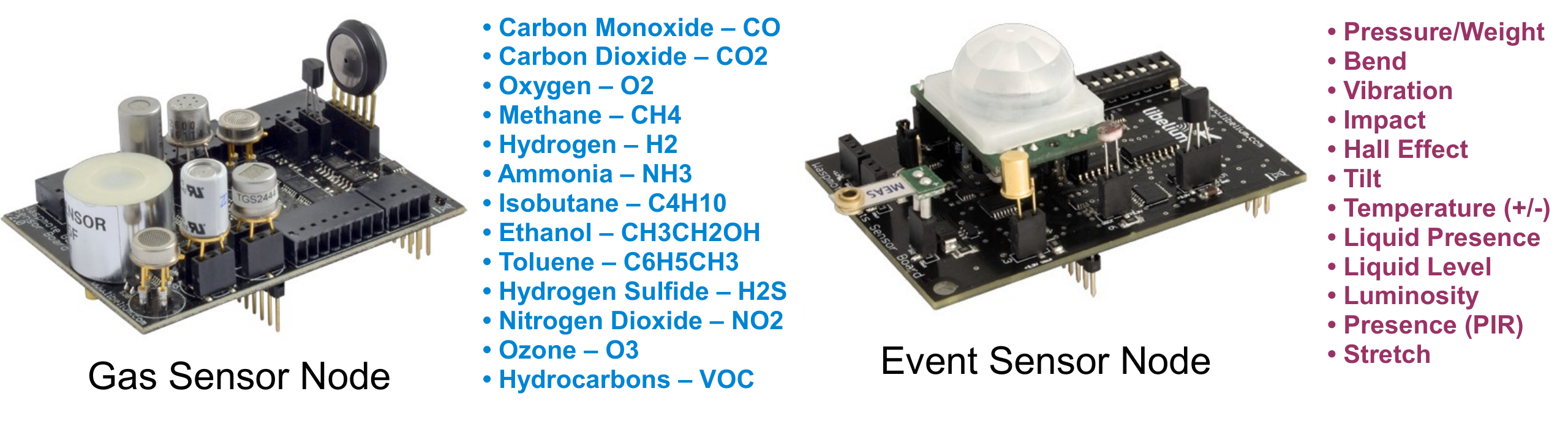}
 \caption{Heterogeneity in term of sensing/measurement capabilities of sensor nodes}
 \label{Figure:Heterogeneity}
\end{figure}

\begin{figure}[h]
 \centering
 \includegraphics[scale=0.40]{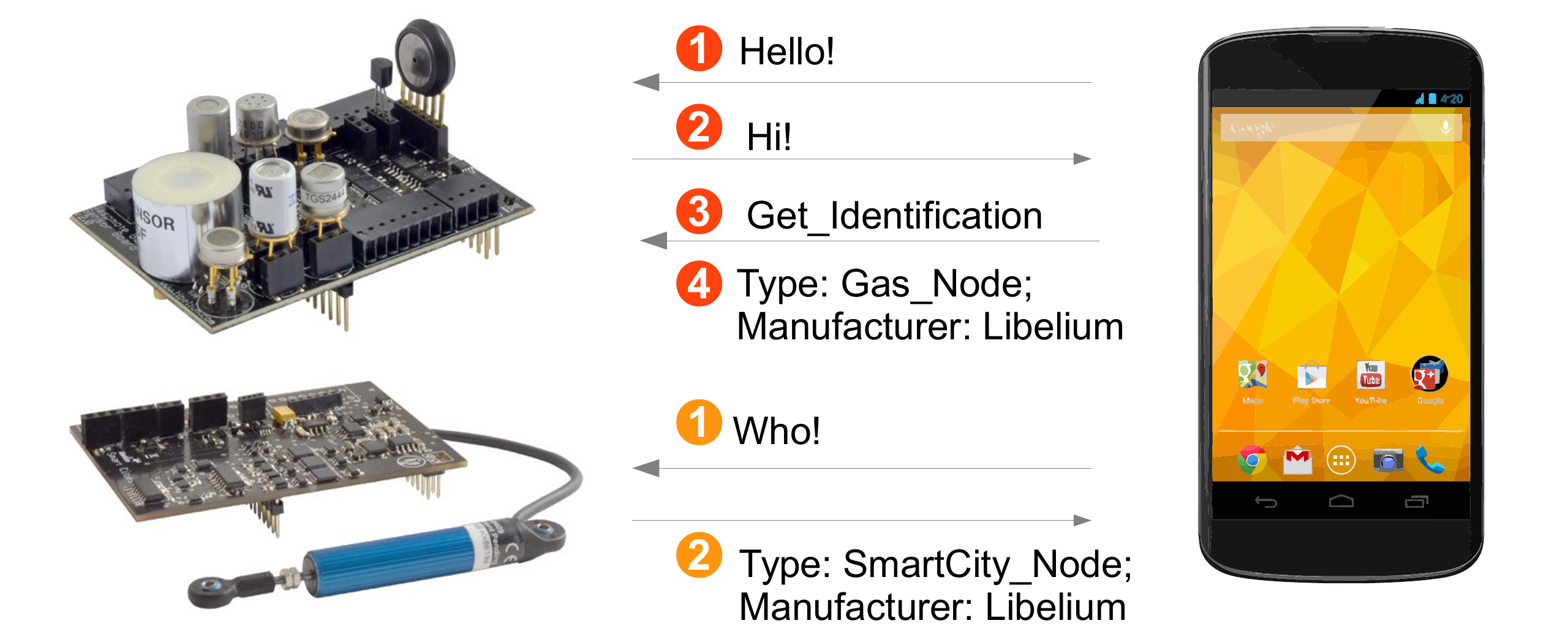}
 \caption{Heterogeneity in term of communication and message/command passing sequences. Some sensors may need only a few  message/command passes and others may require more. The messages/commands understood by each sensor may also vary.}
 \label{Figure:Communication_Sequence}
\end{figure}

\textbf{2) Heterogeneity:} This factor can be interpreted in different perspectives. (1) Heterogeneity in terms of the communication technologies used by the sensors, as presented in Table \ref{Table:Wireless Technology Comparison}. (2) Heterogeneity in terms of measurement capabilities, as presented in Figure \ref{Figure:Heterogeneity} (e.g. temperature, humidity, motion, pressure). (3) The types of data (e.g. numerical (small in size), audio, video (large in size)) generated by the sensors are also heterogeneous. (4) The communication sequences and security mechanisms used by different sensors are also heterogeneous (e.g. exact messages/commands and the sequence that needs to be followed to successfully communicate with a given sensor). As illustrated in Figure \ref{Figure:Communication_Sequence}, some sensors may need only a few command passes and others may require more. Further, the messages/commands understood by each sensor may also vary. These differences make the sensor configuration process challenging. An ideal sensor configuration approach that is designed for the IoT paradigm should be able to handle such heterogeneity. It should also be scalable and should provide support for new sensors as they come to the market.

\begin{table}[t]
\footnotesize
\centering
\caption{Heterogeneity in term of Wireless Communication Technology}
\begin{tabular}{l p{2cm} p{1.8cm} p{1.8cm} p{1.8cm} p{1.8cm}}
\hline 
 & \textbf{ZigBee} & \textbf{GPRS-GSM} & \textbf{WiFi} & \textbf{Bluetooth} \\ \hline \hline 
Standard & 802.15.4 &  & 802.11b & 802.15.1 \\ 
System Resources & 4-32KB & 16MB+ & 1MB+ & 250KM+ \\ 
Batterylife (days) & 100-1000+ & 1-7 & 0.5-5 & 1-7 \\ 
Network Size (nodes) & $2^{64}$ & 1 & 32 & 7 \\ 
Bandwidth (KB/s) & 20-250 & 64-128+ & 11000 & 720 \\ 
Transmission \newline Range (meters) & 1-100+ & 1000 & 1-100 & 1-10+ \\ 
Success Metrics & Reliability, power, cost & Reach, quality &  flexibility, Speed & Convenience, cost \\ \hline
\end{tabular}
\label{Table:Wireless Technology Comparison}
\vspace{-6pt}
\end{table}

\textbf{3) Scheduling, Sampling Rate, and Network Communication:} The sampling rate defines the frequency with which sensors need to generate data (i.e. sense the phenomenon) (e.g. sense temperature every 10 seconds). Deciding the ideal (e.g. balance between user requirement and energy consumption) sampling rate can be a very complex task and has a strong relationship with \textbf{\textit{6) Context}} (see below). The schedule defines the timetable for sensing and data transmission (e.g. sense the temperature only between 8am and 5pm on weekdays). Network communication defines the frequency of data transmission (e.g. send data to the cloud-based IoT platform every 60 seconds). Designing efficient sampling and scheduling strategies and configuring the sensors accordingly is challenging. Specifically, standards need to be developed in order to define schedules that can be used across different types of sensor devices.

\textbf{4) Data Acquisition:} Such methods can be divided into two categories: based on responsibility and based on frequency  \cite{ZMP007}. There are two methods that can be used to acquire data from a sensor based on responsibility: push (e.g. the cloud requests data from a sensor and the sensor responds with data) and pull (e.g. the sensor pushes data to the cloud without continuous explicit cloud requests). Further, based on frequency, there are two data acquisition methods: instant (e.g. send data to the cloud when a predefined event occurs) and interval (e.g. send data to the cloud periodically). Pros, cons, and applicabilities of these different approaches are discussed in \cite{ZMP007}. Using the appropriate data acquisition method based on context information is essential to ensure efficiency.

\begin{figure}[h]
 \centering
 \includegraphics[scale=0.40]{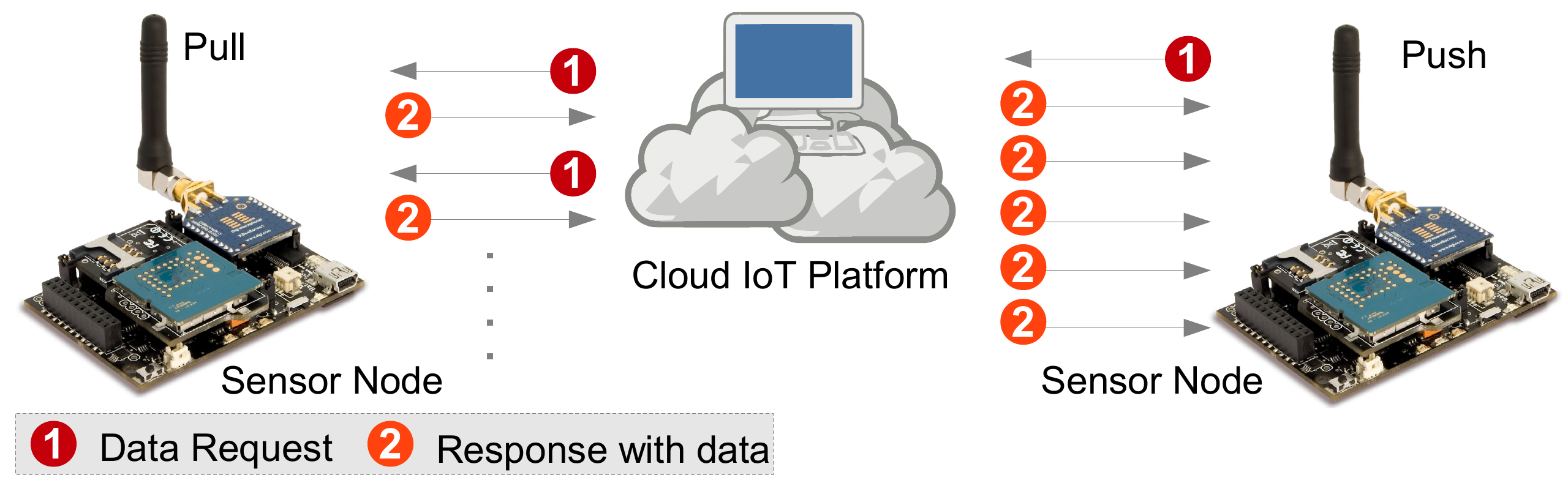}
 \caption{Data can be retrieved from a sensor using both push (right side) and pull (left side) communication methods. Each method has its own advantages and disadvantages which make them suitable for different situations.}
 \label{Figure:Push_and_Pull}
\end{figure}

\textbf{5) Dynamicity:} This means the frequency of changing positions / appearing / disappearing of the sensors at a given location.  IoT envisions that most of the objects we use in everyday lives will have sensors attached to them in the future. Ideally, we need to connect and configure these sensors to software platforms in order to analyse the data they generate and so understand the environment better. We have observed several domains and broadly identified different levels of dynamicity based on mobility\footnote{It is important to note that the same object can be classified at different levels depending on the context. Further, there is no clear definition to classify objects into different levels of dynamicity. However, our categorization allows us to understand the differences in dynamicity.}. Sensors that move/ appear/ disappear at a higher frequency (e.g. RFID and other low-level, low-quality, less reliable, cheap sensors that will be attached to consumables such as stationery, food packaging, etc.) can be classified as highly dynamic. Sensors embedded and fitted into permanent structures (such as buildings and air conditioning systems) can be classified as less dynamic. An ideal sensor configuration platform should be able to efficiently and continuously discover and re-configure sensors in order to cope with high dynamicity.

\textbf{6) Context:} Context information plays a critical role in sensor configuration in the IoT. The objective of collecting sensor data is to understand the environment better by fusing and reasoning  them. In order to accomplish this task, sensor data needs to be collected in a timely and location-sensitive manner. Each sensor needs to be configured by considering context information. Let us consider a scenario related to smart agriculture to understand why context matters in sensor configuration. \textit{Severe frosts and heat events can have a devastating effect on crops. Flowering time is critical for cereal crops and a frost event could damage the flowering mechanism of the plant. However, the ideal sampling rate could vary depending on both the season of the year and the time of day. For example, a higher sampling rate is necessary during the winter and the night. In contrast, lower sampling would be sufficient during summer and daytime. On the other hand, some reasoning approaches may require multiple sensor data readings. For example, a frost event can be detected by fusing air temperature, soil temperature, and humidity data. However, if the air temperature sensor stops sensing due to a malfunction, there is no value in sensing humidity, because frost events cannot be detected without temperature. In such circumstances, configuring the humidity sensor to sleep is ideal until the temperature sensor is replaced and starts sensing again}. Such intelligent (re-)configuration can save energy by eliminating  ineffectual  sensing and network communication.

\vspace{-6pt}

\section{Related Work}
\label{sec:Related_Work}

In this section, we review some of the state-of-the-art solutions developed by the research community, as well as commercial business entities. Our review covers both mature and immature solutions proposed by start-up initiatives as well as large-scale projects. Our proposed CADDOT model as well as the \textit{SmartLink} tool help to overcome some of the weaknesses in the existing solutions.

There are commercial solutions available in the market that have been developed by start-up IoT companies \cite{P596} and the research divisions of leading corporations. These solutions are either still under development or have completed only limited deployments in specialized environments (e.g. demos). We discuss some of the selected solutions based on their popularity. \textit{Ninja Blocks} (ninjablocks.com), \textit{Smart-Things} (smartthings.com), and \textit{Twine} (supermechanical.com) are commercial products that aim at building smart environments \cite{P596}. They use their own standards and protocols (open or closed) to communicate between their own software systems and sensor hardware components. The hardware sensors they use in their solutions can only be discovered by their own software systems. In contrast, our pluggable architecture can accommodate virtually any sensor. Further, our proposed model can facilitate different domains (e.g. indoor, outdoor) using different communication protocols and sequences. 

In addition, the CADDOT model can facilitate very high dynamicity and mobility. \textit{HomeOS}  \cite{P597} is a home automation operating system that simplifies the process of connecting devices together. Similar to our plugin architecture, \textit{HomeOS} is based on applications and drivers which are expected to be distributed via an on-line store called \textit{HomeStore} in the future. However, \textit{HomeOS} does not perform additional configuration tasks (e.g. scheduling, sampling rate, communication frequency) depending on the user requirements and context information. Further, our objective is to develop a model that can accommodate a wider range of domains by providing multiple alternative mechanisms, as discussed in Section \ref{sec:Design_Decisions}. Hu et al. \cite{P339} have proposed a sensor configuration mechanism that uses the information store in TEDS \cite{P258} and SensorML \cite{P256} specifications. Due to the unavailability and unpopularity of TEDS among sensor manufacturers, we simulate  TEDS using standard communication message formats, as explained in Section \ref{sec:Design_Decisions}.

Actinium \cite{P636} is a RESTful runtime container that provides Web-like scripting for low-end devices through a cloud. It encapsulates a given sensor device using a container that handles the communication between the sensor device and the software system by offering a set of standard interfaces for sensor configuration and life-cycle management. The Constrained Application Protocol (CoAP),a software protocol intended to be used in very simple electronics devices that allows them to communicate interactively over the Internet, has been used for communication. Pereira et al. \cite{P641} have also used CoAP and it provides a request/response interaction model between application end-points. It also supports built-in discovery of services and resources. However, for discovery to work, both the client (e.g. a sensor) and the server (e.g. the IoT platform) should support CoAP. However, most of the sensor manufacturers do not provide native support for such protocols. \textit{Dynamix} \cite{P627} is a plug-and-play context framework for Android. \textit{Dynamix} automatically discovers, downloads, and installs the plugins needed for a given context sensing task. \textit{Dynamix} is a stand-alone application and it tries to understand new environments using pluggable context discovery and reasoning mechanisms. Context discovery is the main functionality in \textit{Dynamix}. In contrast, our solution is focused on dynamic discovery and configuration of \things in order to support a sensing as a service model in the IoT domain. We employ a pluggable architecture which is similar to the approach used in \textit{Dynamix}, in order to increase the scalability  and rapid extension development by third party developers. The Electronic Product Code (EPC)  \cite{P110} is designed as a universal identifier that provides a unique identity for every physical object anywhere in the world. EPC is supported by the CADDOT model as one way of identifying a given sensor. Sensor integration using IPv6 in building automation systems is discussed in \cite{P656}. Cubo et al. [12] have used a Device Profile for Web Services\footnote{http://docs.oasis-open.org/ws-dd/ns/dpws/2009/01} (DPWS) to encapsulate both devices and services. DPWS defines a minimal set of implementation constraints to enable secure web service messaging, discovery, description, and eventing on resource-constrained devices. However, discovery is only possible if both ends (client and server) are DPWS-enabled.

\section{Overview of the CADDOT Model}
\label{sec:Architectural_Design}

Previously, we identified several major factors that need to be considered when developing an ideal sensor configuration model for the IoT. This section presents a detailed explanation of our proposed solution: Context-aware Dynamic Discovery of Things (CADDOT). Figure 5 illustrates the main phases of the proposed model.

\textbf{Phases in CADDOT model:} The proposed model consists of eight phases: \textit{detect, extract, identify, find, retrieve, register, reason,} and \textit{configure}.  Some of the tasks mentioned in the model are performed by the \textit{SmartLink} tool and other tasks are performed by the cloud middleware. Some tasks are performed collectively by both \textit{SmartLink} and the cloud.

\begin{figure}[h]
 \centering
 \includegraphics[scale=0.98]{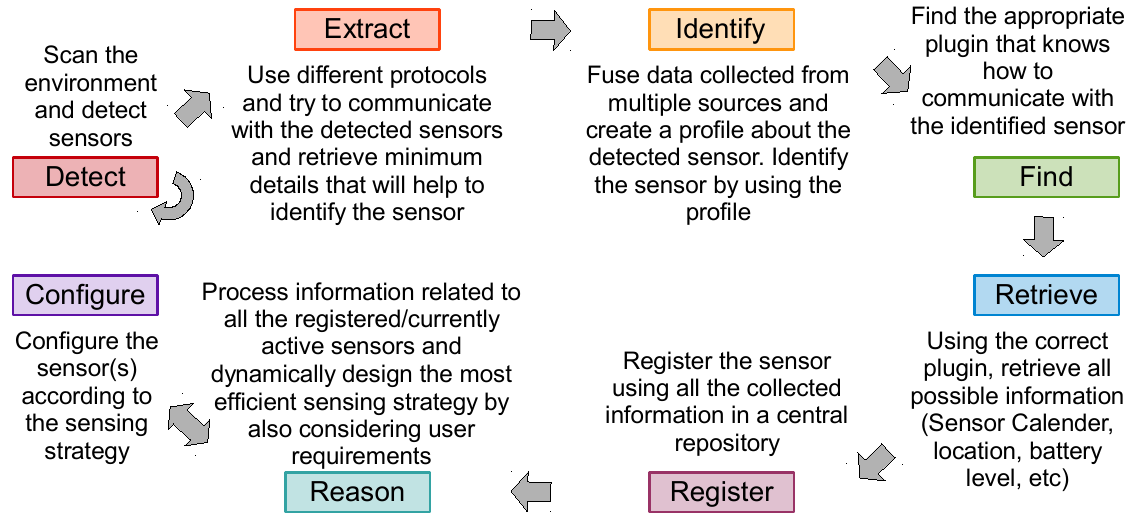}
 \caption{Context-aware Dynamic Discovery of Things (CADDOT) model for configuration of things in the IoT paradigm consists of eight phases.}
 \label{Figure:Model}
\end{figure}

\textit{\textbf{1) Detect:}} Sensors are configured to actively seek open wireless access points (WiFi or Bluetooth) to which they can be connected without any authorization, because in this phase sensors do not have any authentication details. Sensors will receive the authentication details in phase \textbf{phase 8)}. As a result, in this phase sensors are unable to connect to an available secured network. The mobile device that \textit{SmartLink} is installed in becomes an open wireless access point (hotspot) so the sensors can connect to it. However, it is important to note that there are different application strategies that \textit{SmartLink} can use to execute the CADDOT model, as discussed in Section \ref{sec:Design_Decisions}.

\textit{\textbf{2) Extract:}} In this phase, \textit{SmartLink} extracts information from the sensor detected in the previous phase. Each sensor may be designed to respond to different message-passing sequences, as illustrated in Figure \ref{Figure:Communication_Sequence}, depending on the sensor manufacturer and the sensor program developer. Even though the sensors and the \textit{SmartLink} may use the same communication technology/ protocol (e.g. TCP, UDP, Bluetooth), the exact communication sequence can vary from one sensor to another. Therefore, it is hard to find the specific message-passing sequence that each sensor follows. To address this challenge, we propose that every sensor will respond to a common message during the communication initiation process. Alternatively, CADDOT can support multiple initiation messages (extraction mechanisms). However, such alternative approaches will increase the time taken to extract a minimum set of information from a given sensor due to multiple communication attempts that need to be carried out until a sensor successfully responds. For example, \textit{SmartLink} broadcasts a message [WHO], as illustrated in (\textit{C1}) in Figure \ref{Figure:Message_Formats}, where the sensors are expected  to respond by providing a minimum amount of information about themselves, such as a sensor's unique identification number, model number / name, and manufacturer. This is similar to the TEDS mechanism discussed in \cite{P339}. It is important to note that we propose this [WHO] constraint only for minimum information extraction. Once the sensor is identified, subsequent communications and heterogeneity of message-passing sequences are handled by matching plugins.

\textit{\textbf{3) Identify:}} \textit{SmartLink} sends all the information extracted from the newly detected sensor to the cloud. Cloud-based IoT middleware queries its data stores using the extracted information and identifies the complete profile of the sensor. The descriptions of the sensors are modelled in an ontology\footnote{This is an extended version of an SSN ontology (www.w3.org/2005/Incubator/ssn/ssnx/ssn). The detailed description of our extended ontology is out of the scope of this chapter.}.

\textit{\textbf{4) Find:}} Once the cloud identifies the sensor uniquely, this information is used to find a matching plugin (also called drivers) which knows how to communicate with a compatible sensor at full capacity. The IoT middleware pushes the plugin to \textit{SmartLink} where it is installed\footnote{In practice, the IoT middleware sends a request to the application store (e.g. Google Play). The application store pushes the plugin to the \textit{SmartLink} autonomously via the Internet.}.

\textit{\textbf{5) Retrieve:}} Now, \textit{SmartLink} knows how to communicate with the detected sensor at full capacity with the help of the newly downloaded plugin. Next, \textit{SmartLink} retrieves the complete set of information that the sensor can provide (e.g. configuration details such as schedules, sampling rates, data structures /types generated by the sensor, etc.). Further, \textit{SmartLink} may communicate with other available sources (e.g. databases, web services) to retrieve additional information related to the sensor.

\textit{\textbf{6) Register:}} Once all the information about a given sensor has been collected, registration takes place in the cloud. The sensor descriptions are modelled according to the semantic sensor network ontology (SSNO) \cite{P626}. This allows semantic querying and reasoning at a later stage to perform operations such as sensor search \cite{ZMP006}. Some of the performance evaluation related to the SSN ontology and semantic querying is presented in \cite{ZMP011}.

\textit{\textbf{7) Reason:}} This phase plays a significant role in the sensor configuration process. It designs an efficient sensing strategy. Reasoning takes place in a distributed manner. The cloud IoT middleware retrieves data from a large number of sensors and identifies their availabilities and capabilities. Further, it considers context information in order to design an optimized strategy. Context-aware reasoning is performed by IoT middleware on the cloud. However, the technical details related to this reasoning process are out of the scope of this chapter. At the end of this phase, a comprehensive plan (i.e. sensing schedule) for each individual sensor is designed.

\textit{\textbf{8) Configure:}} Sensors as well as cloud-based IoT software systems are configured based on the strategy designed in the previous phase. Schedules, communication frequency, and sampling rates that are custom-designed for each sensor are pushed into the individual sensors. The connections between sensors and the cloud-based IoT software system are established through direct wireless communication or through intermediate devices such as MOSDEN \cite{ZMP009} so the cloud can retrieve data from sensors. The configuration details (e.g. IP address, port, authentication) required to accomplish the above task are also provided to the sensor.

\section{Design Decisions and Applications}
\label{sec:Design_Decisions}

We made a number of design decisions during the development of the CADDOT model. These decisions address the challenges we highlighted in earlier sections.

\begin{figure}[t]
 \centering
 \includegraphics[scale=0.4]{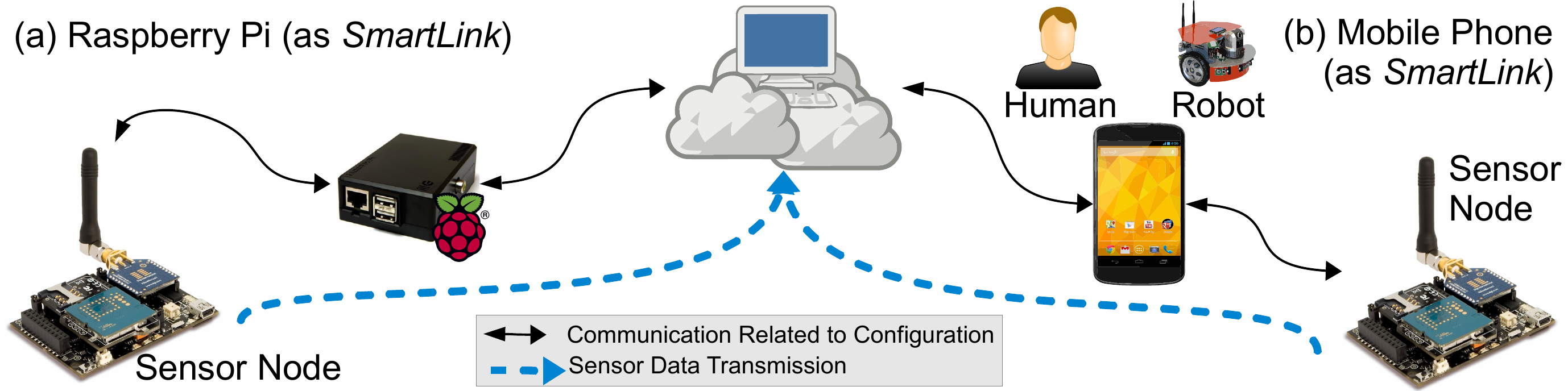}
 \caption{Application strategies of CADDOT model and \textit{SmartLink} tool. (a) usage of static \textit{SmartLink} (b) usage of mobile \textit{SmartLink}.}
 \label{Figure:Usage_Pattern}
\end{figure}

\textbf{Security Concerns and Application Strategies:} There are different ways to employ our proposed model CADDOT as well as the tool \textit{SmartLink} in real world deployments. Figure \ref{Figure:Usage_Pattern} illustrates two different application strategies. It is important to note that neither our model nor the software tool is limited to a specific device or platform. In this paper, we conduct the experimentations on an Android-based mobile phone, as detailed in Section  \ref{sec:Implementation}. In strategy (a), a Raspberry Pi (raspberrypi.org) is acting as the \textit{SmartLink}  tool. This strategy is mostly suitable for smart home and office environments where WiFi is available. Raspberry Pi continuously performs the discovery and configuration process, as explained in Section  \ref{sec:Architectural_Design}. Finally, Raspberry Pi provides the authentication details to the sensor which is connected to the secure home/office WiFi network. The sensor is expected to send data to the processing server (local or on cloud) directly over the secured WiFi network. In this strategy, \textit{SmartLink} is in static mode. Therefore, several \textit{SmartLink} installed Raspberry Pi devices may be required to cover a building.  However, this strategy can handle a high level of dynamicity.

The \textit{{strategy (b)} }is  more suitable for situations where WiFi is not available or less dynamic. Smart agriculture can be considered as an example. In this scenario, sensors are deployed over a large geographical area (e.g. Phenonet \cite{P412}). Mobile robots\footnote{In small agricultural fields, farmers themselves can carry the \textit{SmartLink} over the field.} (tractors or similar vehicles) with a \textit{SmartLink} tool attached to them can be used to discover and configure sensors. \textit{SmartLink} can then help to establish the communication between sensors and sinks. The permanent sinks used in the agricultural fields are usually low-level sinks (such as Messhablium \cite{P595}). Such sinks cannot perform sensor discovery or configuration in comparison to SmartLink. Such sinks are designed to collect data from sensors and upload to the cloud via 3G.

Many more different strategies can be built by incorporating the different characteristics pointed out in the above two strategies. This shows the extensibility of our solution. For example, Raspberry Pi, which we suggested for use as a \textit{SmartLink} in strategy a), can be replaced by corporate mobile phones. So, without bothering the owner, corporate mobile phones can silently perform the work of a \textit{SmartLink}.

\begin{figure}[t!]
 \centering
 \includegraphics[scale=0.55]{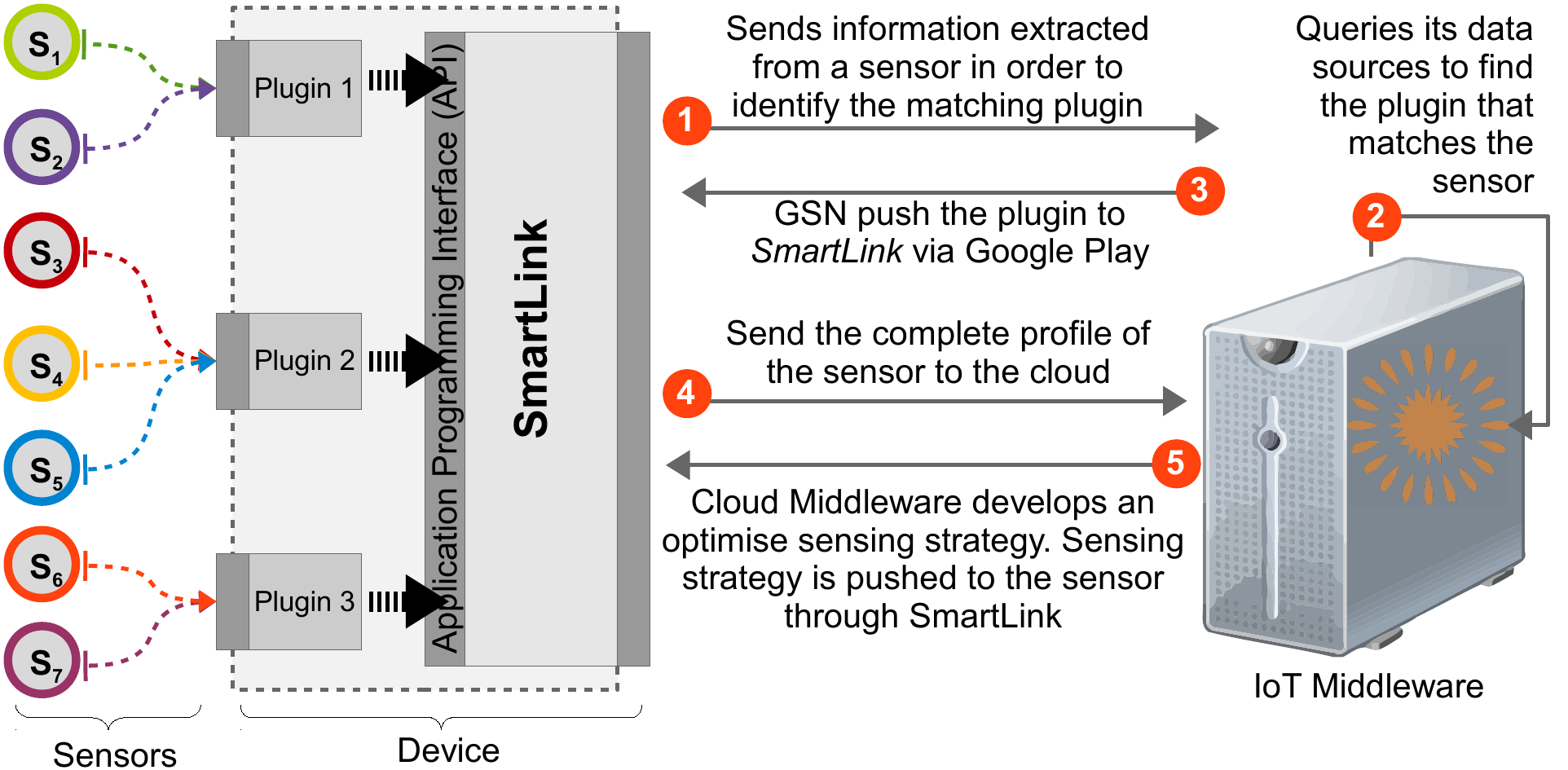}
 \caption{System architecture of the CADDOT model which consists of three main components: sensors, \textit{SmartLink} tool, and the cloud middleware. Interactions are numbered in order.}
 \label{Figure:System}	
\vspace{-0.33cm}	
\end{figure}

\textbf{System Architecture:} The CADDOT model consists of three main components: sensors, a mobile device (i.e. \textit{SmartLink}), and the cloud middleware. All three components need to work collectively in order to perform sensor discovery and configuration successfully. Figure  \ref{Figure:System} illustrates the interactions between the three components. The phases we explained earlier relating to the CADDOT model in Figure \ref{Figure:Model} can be seen in Figure \ref{Figure:System} as well. As we mentioned before, \textit{SmartLink} is based on a plugin architecture. The core \textit{SmartLink} application cannot directly communicate with a given sensor. A plugin needs to act as a mediator between the sensor and the \textit{SmartLink} core application, as illustrated in Figure \ref{Figure:System}. The task of the mediator is to translate the commands back and forth. This means that in order to configure a specific sensor, the \textit{SmartLink} core application needs to employ a plugin that is compatible with both the \textit{SmartLink} application itself and the given sensor. We discuss this matter in the programming perspective later in this section.

\textbf{Sensor-level Program Design:}
One of the most important components in the CADDOT model is the sensor. Sensors can be programmed in different ways. In this chapter, we propose a program design that supports all the functional requirements identified in Section \ref{sec:Functional_Requirements}. The program we propose may not be the only way to support these requirements. Further, we do not intend to restrict developers to one single sensor-level program design. Instead, our objective is to demonstrate one successful way to program a sensor in such a way that it allows sensors to be re-configured at runtime (i.e. after deployment) depending on the requirements that arise later. Developers are encouraged to explore more efficient program designs. However, in order to allow \textit{SmartLink} to communicate with a sensor which runs different program designs, developers need to develop a plugin that performs the command translations. We explain the translation process using both sensor-level program code as well as plugin code later in this section. First, we illustrate the simplest sensor-level program that can be designed to perform the task of sensing and transmitting data to the cloud in Figure 8. We refer to this program design as \textit{SPD} (Simple Program Design) hereafter. The basic structure of a sensor-level program is explained in \cite{P595}.

\begin{figure}[t]
 \centering
 \includegraphics[scale=1.0]{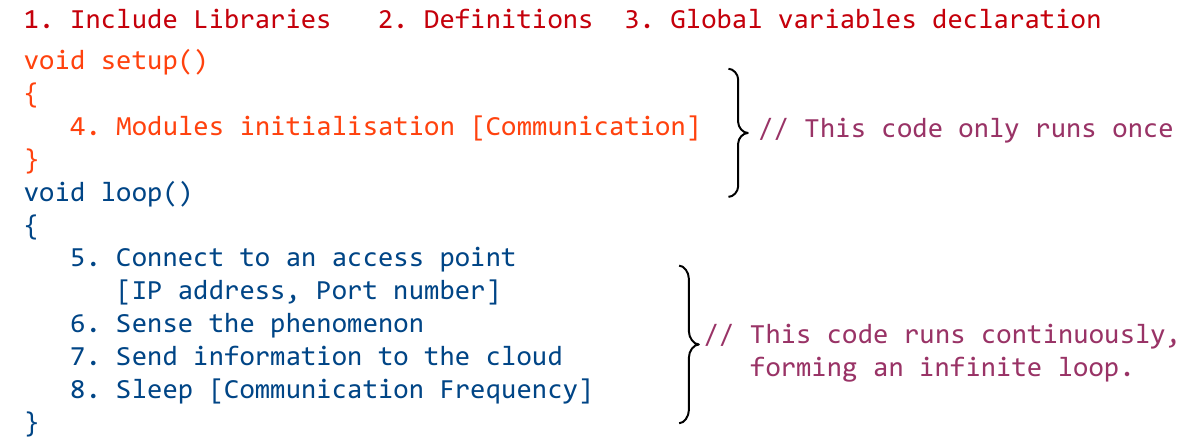}
 \caption{A simple sensor-level program design (SPD) that sends and transmits data to the cloud. It does not support dynamic discovery and configuration.}
 \label{Figure:Simple_Sensor_Program}
\end{figure}

\begin{figure}[h]
 \centering
 \includegraphics[scale=0.90]{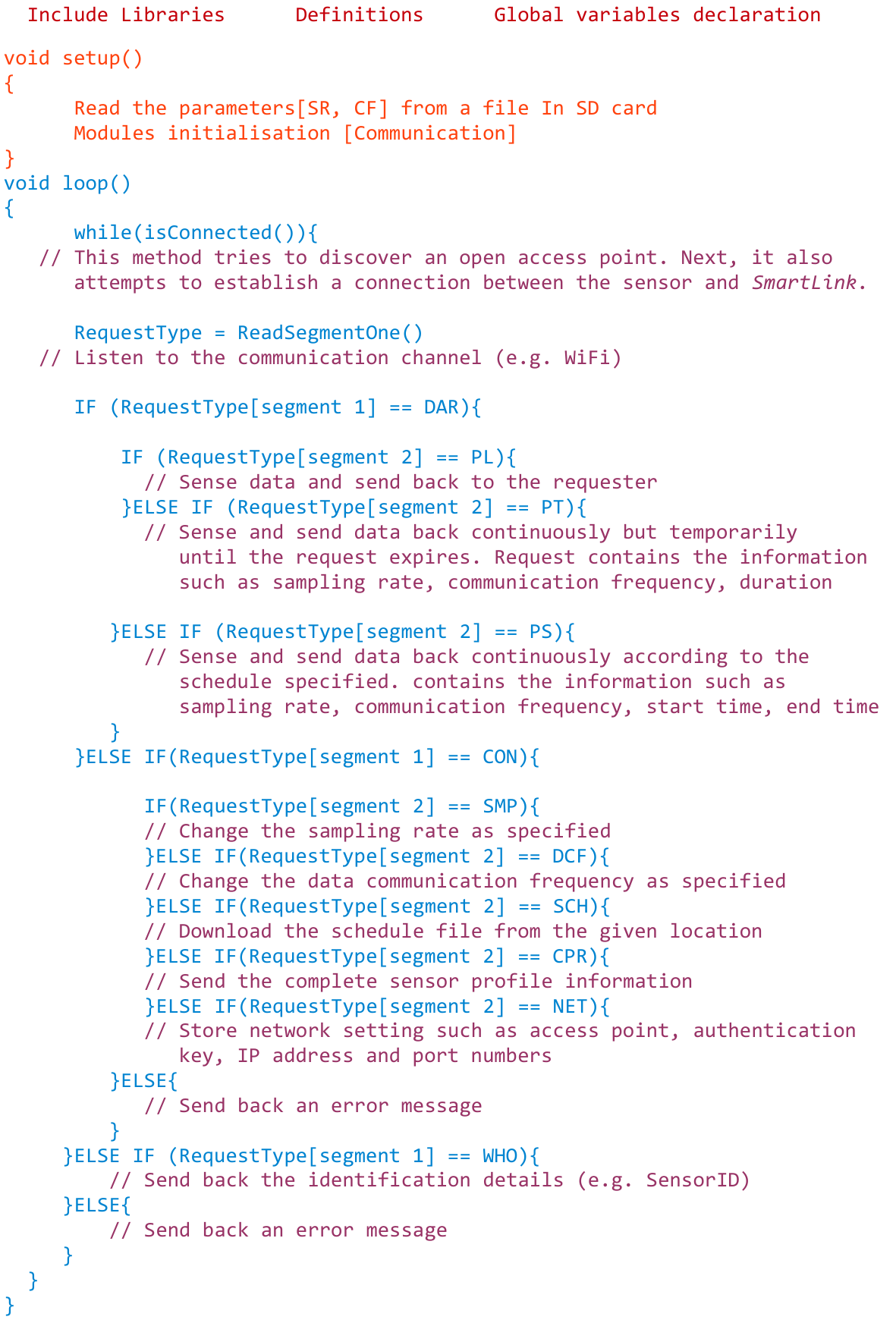}
 \caption{A configurable sensor-level program design (CPD) that supports dynamic discovery and configuration after deployment at runtime.}
 \label{Figure:Configurable_Sensor_Program}
\vspace{-18pt}
\end{figure}

\begin{figure}[h]
 \centering
 \includegraphics[scale=0.76]{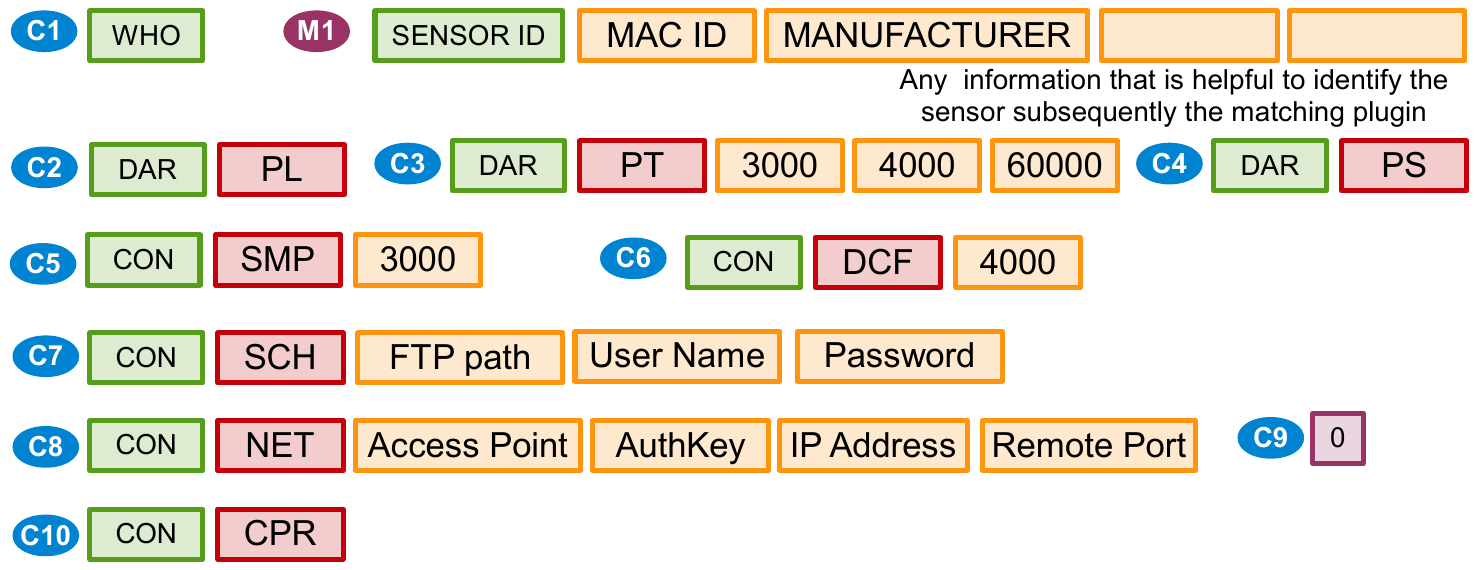}
 \caption{Command formats used to perform sensor configuration.}
 \label{Figure:Message_Formats}
\vspace{-8pt}
\end{figure}

The main problem in this program design is that there is no way to configure (i.e. sampling rate, communication frequency, data acquisition method) the sensor after deployment other than by re-programming (e.g. Over the Air Programming). However, such re-programming approaches are complex, labour-intensive and time consuming. In Figure  \ref{Figure:Configurable_Sensor_Program}, we designed a sensor-level program that supports a comprehensive set of configuration functionalities. We refer to this design as \textit{CPD} (Configurable Program Design) hereafter. In order to standardize the communication, we also defined a number of command formats. However, these messaging formats do not need to be followed by the developers as long as they share common standardised command formats between their own sensor-level program and the corresponding plugin. Different command formats used to accomplish different tasks in our approach are illustrated in Figure \ref{Figure:Message_Formats}. In comparison to \textit{SPD}, \textit{CPD} provides more configuration functionalities. With the help of the command formats illustrated in Figure \ref{Figure:Message_Formats}, \textit{SmartLink} can configure a given sensor at any time.

Each command comprises several different segments, as depicted in Figure \ref{Figure:Message_Formats}. The first segment denotes whether the command is related to configuration or a data request. In our approach, [CON] denotes configuration and [DAR] denotes a data request. The CPD is designed to forward the command appropriately through IF-ELSE branches. The CPD accepts five different types of commands under the [CON] branch. Commands are classified based on the second segment. The following list summarises these commands. The first segment of every command contains only three letters which makes it easy to process. The commands can be sent using frames\footnote{http://www.libelium.com/uploads/2013/02/data\_frame\_guide.pdf}  or plain strings.

\begin{itemize}
\item \textbf{C1:} This command has only one segment.  This segment always contains three letters [WHO]. This command is sent by \textit{SmartLink} to a sensor. To support CADDOT, every sensor should be able to handle command C1. Then the sensor needs to respond with message \textbf{M1}. This is the only constraint that the sensor-level program developers are required to adhere to.

\item \textbf{M1:} This message is sent by the sensor to \textit{SmartLink} in response to C1. M1 contains information that helps to identify the sensor in \textit{key-value pair} format. The information contained in this message is sent to the cloud IoT platform, as explained in phase (4) in the CADDOT model illustrated in Figure \ref{Figure:Model}. Detailed explanation of this message is out of the scope of this chapter.

\item \textbf{C2:}  This command consists of two segments. The first segment [DAR] denotes that this is a data request. The second segment [PL] denotes that the command is a pull request which the sensor is expected to respond to with sensors data once.

\item \textbf{C3:}  This command consists of five segments. The first segment [DAR] denotes that this is a data request. The second segment [PS] denotes that the sensor is expected to push data according to the information provided in the rest of the segments. The third segment specifies the sample rate and the fourth segment specifies the data communication frequency rate. The final segment specifies the duration for which the sensor needs to push data to the cloud.

\item \textbf{C4:} This command consists of two segments. The first segment [DAR] denotes that this is a data request. The second segment [PS] denotes that the sensor is expected to perform sensing and data transmitting tasks according to a sensing schedule specified in the sensing schedule file. It is expected to push data to the cloud.

\item \textbf{C5:} This command consists of three segments. The first segment [CON] denotes that this is a configuration command. The second segment [SMP] denotes that this command configures the sampling rate. The third segment holds the actual sampling rate value that the sensor needs to sense in the future.

\item \textbf{C6:} This command consists of three segments. The first segment [CON] denotes that this is a configuration command. The second segment [DCF] denotes that this command configures the data communication frequency. The third segment holds the actual data communication frequency rate value that the sensor needs to transmit data to the cloud in the future.

\item \textbf{C7:} This command consists of five segments. The first segment [CON] denotes that this is a configuration command. The second segment [SCH] denotes that this command configures the sensing schedule. The rest of the segments contain information that is essential (i.e. FTP server path, user name, password) to download a sensing schedule file from an FTP server, as depicted in Figure \ref{Figure:Message_Formats}.

\item \textbf{C8:} This command consists of seven segments. The first segment [CON] denotes that this is a configuration command. The second segment [NET] denotes that this command configures the network settings. The rest of the segments contain the information  that is essential to connect to a secure network (i.e. access point name, authentication key, IP address, remote port) so the sensor can directly communicate with the cloud IoT platform.

\item \textbf{C9:} This command stops the sensor completely and pushes it back to a state where the sensor listens for the next command.

\item \textbf{C10:} This command consists of two segments. The first segment [CON] denotes that this is a configuration command. The second segment [CPR] denotes that the sensor is expected to reply with the complete sensor profile.

\end{itemize}

\textbf{Scalable and Extensible Architecture:} As we mentioned earlier, the reason for employing a plugin architecture is to support scalability and extensibility.  Plugins that are compatible with \textit{SmartLink} can be developed by anyone as long as they follow the basic design principles and techniques explained below. Such a plugin architecture allows us to engage with developer communities and support a variety of different sensors through community-based development. We expect to release our software as free and open source software in the future. We provide the main \textit{SmartLink} application as well as the standard interfaces which developers can use to start to develop their own plugins to support different sensors. We provide sample plugin source code where developers only need to add their code according to the guidelines provided. The plugin architecture  will enable more number of sensors to be supported by \textit{SmartLink} over time. Applications stores (e.g. \textit{ Google Play}) built around the Android ecosystem provide an easy way to share and distribute plugins for  \textit{SmartLink}. The pluggable architecture dramatically reduces the sensor configuration time.

\begin{figure}[h]
 \centering
 \includegraphics[scale=1.0]{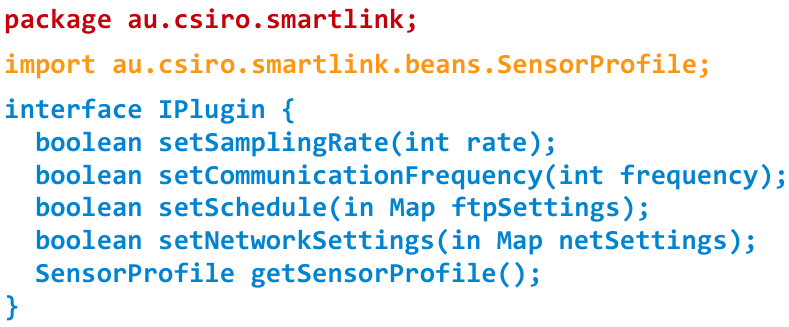}
 \caption{IPlugin written in AIDL (Android Interface Definition Language) that governs the plugin structure. It defines the essential methods that need to be implemented in the plugin class.}
 \label{Figure:AIDL_Code}	
\vspace{-0.33cm}	
\end{figure}

Let us explain how third party developers can develop plugins in such a way that their plugins are compatible with \textit{SmartLink} so that \textit{SmartLink} can use the plugins to configure sensors at runtime when necessary. In plugin development, there are three main components that need to be considered: (1) the plugin interface written in the Android Interface Definition Language (AIDL), (2) the plugin class written in Java, and (3) the plugin definition in the AndroidManifest file. Figure \ref{Figure:AIDL_Code} shows the plugin interface written in AIDL. \textit{IPlugin} is an interface defined in AIDL. Plugin developers should not make any changes in this file. Instead they can use this file to understand how the \textit{SmartLink} plugin architecture works. \textit{IPlugin} is similar to a Java interface.  It defines all the methods that need to be implemented by all the plugin classes.

\begin{figure}[h]
 \centering
 \includegraphics[scale=0.85]{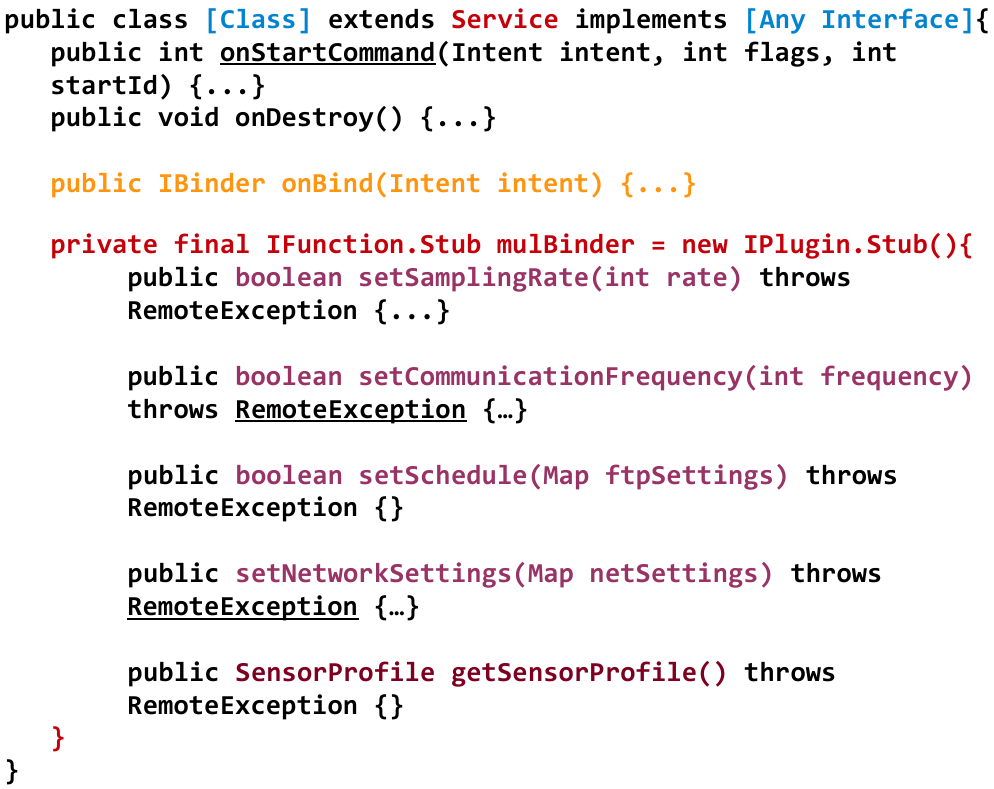}
 \caption{\textit{SmartLink} plugin is an Android service. This is the basic structure of a \textit{SmartLink} plugin. The body of each method needs to be added by the developer based on the sensor-level program design.}
 \label{Figure:Plugin_Code}	
\end{figure}

Figure \ref{Figure:Plugin_Code} presents the basic structure of a \textit{SmartLink} plugin. Each plugin is defined as an Android service. \textit{SmartLink} plugin developers need to implement five methods: \textit{setSamplingRate(int  rate)}, \textit{setCommunicationFrequency(int  frequency)}, \textit{setSchedule(in Map ftpSettings)}, \textit{setNetworkSettings(in Map netSettings)} and \textit{getSensorProfile()}. The methods are briefly explained below.

\begin{itemize}
\item \textit{setSamplingRate(int rate)}:  This method needs to send a command specifying the required sampling rate. For example, in our approach, we defined such a command, \textit{C5}, in Figure \ref{Figure:Message_Formats}.

\item \textit{setCommunicationFrequency(int frequency)}: This method needs to send a command specifying the required communication frequency. For example, in our approach, we defined such a command as \textit{C6} in Figure  \ref{Figure:Message_Formats}.

\item \textit{setSchedule(in Map ftpSettings)}: This method needs to send a command  specifying details (e.g. user-name, password, FTP path) that are required to connect to an FTP server and download  the schedule. For example, in our approach, we defined such a command as, \textit{C7}, in Figure \ref{Figure:Message_Formats}.

\item \textit{setNetworkSettings(in Map netSettings)}:This method sends a command  specifying the details that are required to connect to a secure network  so that direct  communication  between the sensor and the cloud IoT platform  can be established. For example, in our approach, we defined such a command, \textit{C8}, in Figure \ref{Figure:Message_Formats}.

\item \textit{getSensorProfile()}: This method sends a command  to the sensor by asking for profile information.  The sensor is expected to reply by providing  information  such as the data structure it produces, measurement units, and so on. Details of the sensor profiling are out of the scope of this chapter.

\end{itemize}

\begin{figure}[h]
 \centering
 \includegraphics[scale=1.15]{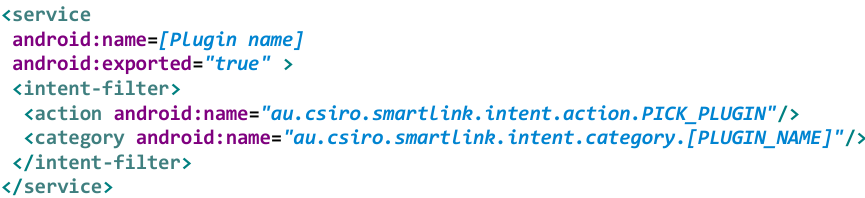}
 \caption{Code snippet of the plugin's \textit{AndroidManifest} file.}
 \label{Figure:AndroidManifest}	
\vspace{-0.23cm}	
\end{figure}

Figure \ref{Figure:AndroidManifest} shows how the plugins need to be defined in the AndroidManifest so that the \textit{SmartLink} application can automatically query and identify them. The Android plugin must have an intent filter which has action name  \textit{\seqsplit{au.csiro.smartlink.intent.action.PICK PLUGIN}}. Developers can provide any category name.

\textbf{Support and Utilize Existing Solutions:} Our model utilizes a few existing solutions. We employed Global Sensor Network  \cite{P022} as the cloud IoT middleware. In CADDOT, GSN performs phases 3, 4, and 7. GSN is a widely used platform in the sensor data processing domain and is used in several European projects, including OpenIoT \cite{P377}. MOSDEN \cite{ZMP009} is middleware that collects sensor data. MOSDEN is ideal for the application strategies we discussed in Section \ref{sec:Design_Decisions} (Figure \ref{Figure:Usage_Pattern}) for use in conjunction with \textit{SmartLink}. \textit{SmartLink} only performs the configuration. Sensor data collection needs to be performed by either cloud IoT middleware or solutions like MOSDEN. The proposed CADDOT model as well as the \textit{SmartLink} tool complement the other solutions proposed by us as well as other researchers. Together, these solutions enable smooth data flow from sensors to the cloud autonomously.

\section{Implementation and Experiment Testbed}
\label{sec:Implementation}

We deployed the \textit{SmartLink} application in a Google Nexus 4 mobile phone (Qualcomm Snapdragon S4 Pro CPU and 2 GB RAM), which runs the Android platform 4.2.2 (Jelly Bean). We deployed 52 sensors on the third floor of the CSIT building (\#108) at the Australian National University. All sensors we employed in our experiment are manufactured by Libelium \cite{P595}. The sensors we used sense a wide variety of environmental phenomena, such as temperature, proximity \& presence, stretch, humidity and so on \cite{P595}. \textit{SmartLink} supports sensor discovery and configuration using both WiFi and Bluetooth. Other communication technologies such as ZigBee  and RFID are supported through Libelium \textit{Expansion Radio Boards} \cite{P595}. In order to simulate the heterogeneity of the sensors (in terms of communication sequence), we programmed each sensor to behave and respond differently. As a result, each sensor can only communicate with a plugin that supports the same communication sequence.

\section{Evaluation of the Prototype}
\label{sec:Evaluation_of the_Prototype}

\begin{figure}[t]
 \centering
 \includegraphics[scale=0.55]{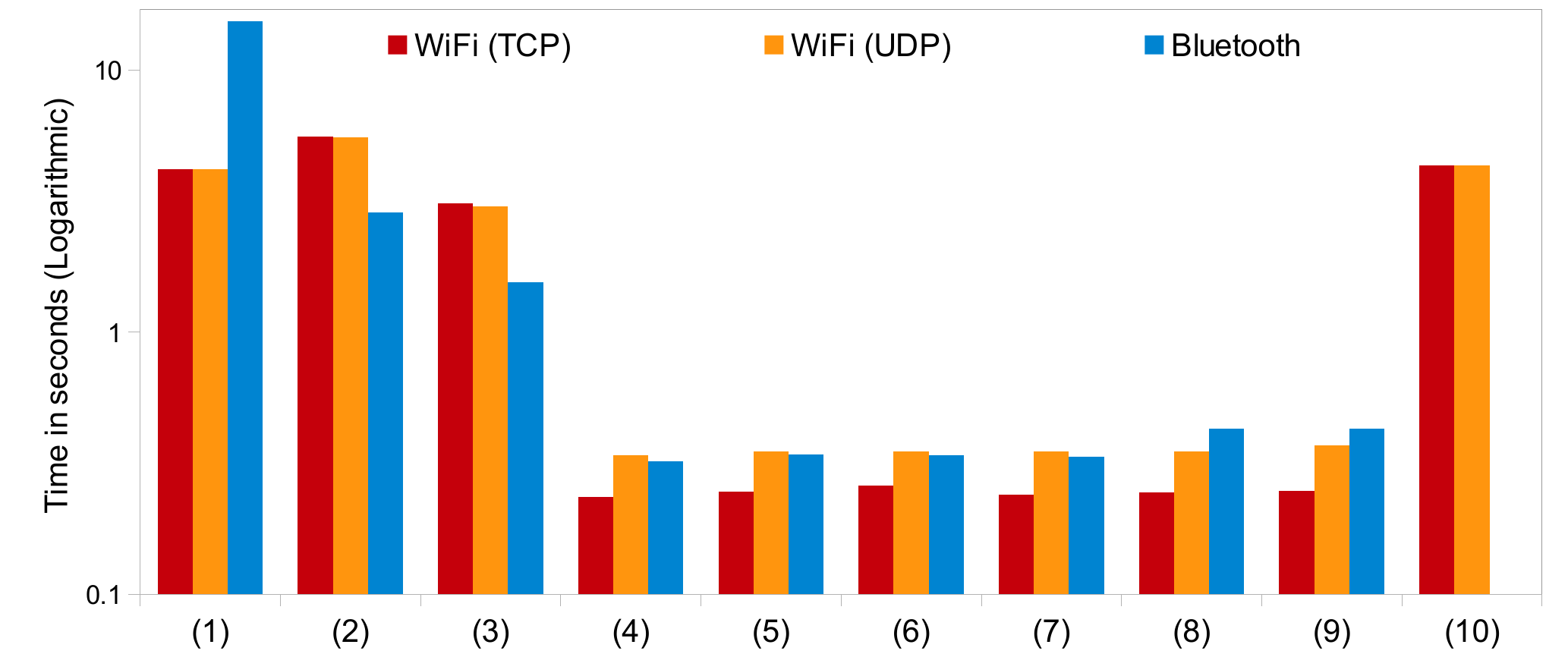}
 \caption{Time taken (y-axis) to discover and configure a sensor step-by-step (x-axis). The experiments were conducted using three  protocols: TCP, UDP, and Bluetooth. }
 \label{Figure:Results1}
\end{figure}

In this section, we explain how we evaluate the proposed CADDOT model  and \textit{SmartLink} tool  using prototype implementations. We identified ten steps performed in the dynamic discovery and sensor configuration process. We measured the average amount of time taken by each of these steps (average of 30 sensor configurations). Figure \ref{Figure:Results1} illustrates the results and the following steps are considered: 
Time taken to (1) set up the sensor, (2) initiate connection between the sensor and \textit{SmartLink}, (3) initiate communication between sensor and \textit{SmartLink}, (4) extract sensor identification information, (5) retrieve the complete profile of the sensor, (6) configure the sampling rate, (7) configure the communication  frequency, (8) configure  the sensing schedule, (9) configure the network and authentication  details (so the sensor can directly connect to the cloud), and (10) connect to the secure network using the provided authentication details.

\textbf{Results:} According to the results, the actual configuration tasks take less that one second. There is a slight variation in completion time in  configuration  step (4) - (9). This is  due to storage access and differences in processing of configuration commands. Sensors takes comparatively longer time to connect to a network as well as to discover and connect to \textit{SmartLink}. Especially, Bluetooth takes much longer to scan for devices in a given environment before it discovers and connects to \textit{SmartLink}. Configuration is slightly faster when using  TCP in comparison to UDP and Bluetooth. This is mainly due to reliability. However, the time differences are negligible. FTP is used to retrieve a scheduling file from a file server. This can take 15-25 seconds depending on the network availability, traffic, and file size. If a sensor cannot access a server via the Internet, a file can be transferred from \textit{SmartLink} to the sensor as typical commands. Sensors generate the scheduling file using the data it receives from \textit{SmartLink}. When using WiFi, a sensor may takes up to 4.5 seconds to connect to a secure network (e.g. WPA2). In contrast, sensors can connect to \textit{SmartLink}'s open access point in less than four seconds.  Despite the protocol we use, sensors take 5 to 15 seconds to boot and setup themselves. The setup stage consists of activities such as reading default configuration from files, and  switching necessary modules and components (communication modules, real-time clock, SD card, sensor broads and so on).

\section{Discussion and Lessons Learned}
\label{sec:Discussion}
In what follows, we discuss major lessons we learned along with limitations. According to our results, it is evident that a single sensor can be configured in less than 12 seconds (i.e. assuming sensors are already booted, which takes an additional 5 to 15 seconds depending on the communication protocol). This is a significant improvement over a manual labour intensive sensor configuration approach. Additionally, \textit{SmartLink} can engage with  number of sensor configuration processes at a given time in parallel. The proposed CPD has not made any negative impact towards the sensing functionality though it supports advance configuration capabilities. The IF-ELSE structure used in CPD makes sure that each request gets to the destination with minimum execution of lines (e.g. `PL' request passes through only two IF conditions). Such execution reduced the impact on sensing tasks while configuration tasks are also supported efficiently. Even though a detailed discussion on data acquisition methods is out of scope, it is important to note that pull, temporary push, and schedule based push add a significant amount of flexibility where each of the techniques is suitable to be used in different circumstances \cite{ZMP007}. The cloud server has the authority to decide which method to be used based on the context information. This increases the efficiency and application scenario where the sensors can be used in sustainable (i.e. in term of energy) manner. Once the initial discovery and configuration of smart things are done, further configuration can be done in more user friendly manner by using techniques such as augmented reality \cite{E1}.


\section{Open Challenges}
\label{sec:Open_Challenges}

In this section, we briefly introduce some of the major open research challenges in the domain that are closely related to this work. We identify four main challenges that provide different research directions.

\textbf{Sensing strategy optimization:} We briefly highlighted the importance of optimizing sensing schedules based on context information in Section \ref{sec:Functional_Requirements}. Sensing strategy development encapsulates a broad set of actions such as deciding the sensing schedule, sampling rate, and network  communication frequency for each sensor. Such a development process needs to consider two main factors: user requirements and availability of sensors. In IoT, there is no single point of control or authority. As a result, different parties are involved in sensor deployments. Such disorganized and uncoordinated deployments can lead to redundant sensor deployment. In order to use the sensor hardware in an optimized manner, sensing strategies need to be developed by considering factors such as sensor capabilities, sensor redundancies (e.g. availability of multiple sensors that are capable of providing similar data), and energy availability. Energy conservation is a key in sustainable IoT infrastructure because the resources constrained  nature of the sensors. We provided such an example in Section 3 related to the agricultural domain. We believe that sensing as a service is a major business model that could drive IoT in the future. In such circumstances, collecting data from all the available sensors has no value. Instead, sensor data should be collected and processed only in response to consumer demand \cite{ZMP008}.

\textbf{Context discovery:} This is an important task where discovered information will be used during a reasoning process (e.g.sensing strategy development). \textit{``Context is any information that can be used to characterise the situation of an entity. An entity is a person, place, or object that is considered relevant to the interaction between a user and an application, including the user and applications themselves''} \cite{P104}. Further discussion on context information and its importance for the IoT is surveyed in \cite{ZMP007}. Context-based reasoning can be used to improve the efficiency of the CADDOT model where a matching plugin can be discovered faster, especially in situations where a perfect match cannot be found. For example, the location of a given sensor\footnote{Location can be represented in many ways: GPS coordinate (e.g. -35.280325, 149.113166), name of a building (e.g. CSIT building  at ANU), name of a city (e.g. Canberra), part of a building  (e.g. living room), floor of a building (e.g. 2nd floor), specific part of a room (e.g. kitchen-top).}, sensors nearby, details of the sensors configured recently, historic data related to sensor availability in a given location, etc. can be fused and reasoned using probabilistic techniques in order to find a matching plugin in an efficient manner. After integrating sensors into cloud-based IoT, the next phase is collecting data from the sensors. Annotating context information to retrieve  sensor data plays a significant role in querying and reasoning them in later stages. Especially, in the sensing as a service model, sensor data consumers may demand such annotation so that they can feed data easily into their own data processing applications for further reasoning and visualization tasks. Some context information can be easily discovered at sensor-level (e.g. battery level, location) and others can be discovered at the cloud-level by fusing multiple raw data items (e.g. activity detection). Such context annotated data help to perform more accurate fusing and reasoning at the  cloud level \cite{E7}.

\begin{figure}[h]
 \includegraphics[scale=.58]{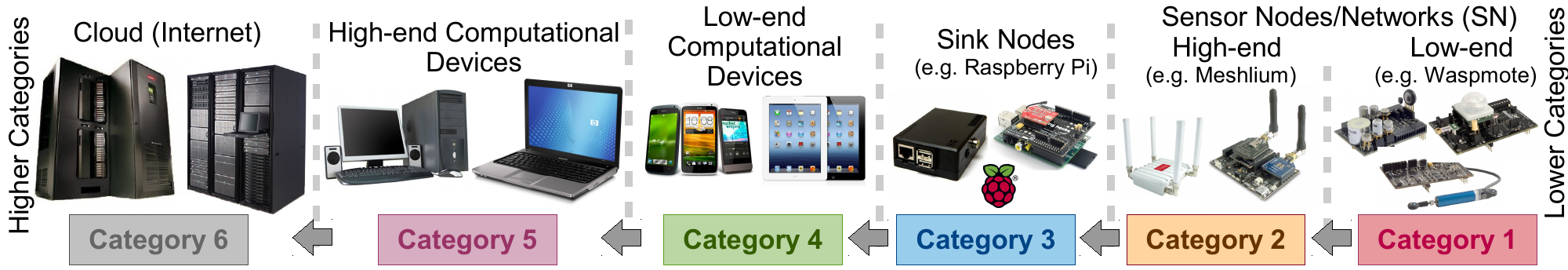}
 \caption{Categorization of IoT devices based on their computational capabilities. The devices belonging to each category have different capabilities in terms of processing, memory, and communication.  They are also different in price, with devices becoming more expensive towards the left. The computational capabilities also increase towards the left.}
 \label{Figure:Layered_Architecture}	
\vspace{-0.20cm}	
\end{figure}

\textbf{Utilization of heterogeneous computational devices:} Even though the IoT envisions billions of \things to be connected to the Internet, it is not possible and practical to connect all of them to the Internet directly. This is mainly due to resource constraints (e.g. network communication capabilities and energy limitations).  Connecting directly to the Internet is expensive in terms of computation, bandwidth use, and hardware costs. Enabling persistent Internet access is challenging and also has a negative impact on miniaturization and energy consumption of the sensors. Due to such difficulties, IoT solutions need to utilize different types of devices with different resource limitations and capabilities. In Figure  \ref{Figure:Layered_Architecture}, we broadly categorise these devices into six categories (also called levels or layers). Devices on the  right side may use low-energy short distance wireless communication protocols to transmit the collected sensor data to the devices on the left. Devices on the left can use long distance communication protocols to transmit the data to the cloud for further processing. However, the more devices we use in smart environments, the more difficult it becomes to detect faults where an entire system could fail \cite{E3}. Providing a unified middleware support across heterogeneity of devices with wider rage of capabilities is an open challenge \cite{E9, E10}.

\textbf{Security and privacy:} In this work, we considered some degree of security as briefly discussed in Section \ref{sec:Design_Decisions}. However, research on security in the IoT is largely unexplored. Security and privacy need to be provided at both sensor-level and cloud-level. It is critical to develop a security model to protect the sensor configuration process, considering questions such as (1) \textit{when to allow reconfiguration of a sensor}, (2) \textit{who has the authority to configure a sensor at a given time}, (3) \textit{how to change ownership of a sensor}, (4) \textit{how to detect sensors with harmful programs installed on them that may cause security threats to a network}. Security and privacy concerns related to the IoT are presented in \cite{P632}. Additionally, security challenges unique to the sensing as a service model are discussed in \cite{ZMP008}.

\section{Conclusions and Outlook}
\label{sec:Conclusions}

In this chapter, we addressed the challenge of integrating sensors into cloud-based IoT platforms through context-aware dynamic discovery and configuration. Traditionally, integration of \things to software solutions is considered a labour-intensive, expensive and time-consuming task that needs to be carried out by  technical experts. Such challenges hinders the non-technical users from adopting IoT to build smart environments. To address this problem, we presented the CADDOT model, an approach that automates the sensor discovery and configuration process in smart environments efficiently and effortlessly by handling key challenges such as a higher number of sensors available, heterogeneity, on-demand sensing schedules, sampling rate, data acquisition methods, and dynamicity. It also encourages non-technical users to adopt IoT solutions with ease by promoting automatic discovery and configuration IoT devices.

 In this work, we supported and evaluated different types of communication technologies (i.e. WiFi and Bluetooth), application strategies, and sensor-level program designs, each of which has their own strengths and weaknesses. We validate the CADDOT model by deploying it in an office environment. As CADDOT required minimum user involvement and technical expertise, it significantly reduces the time and cost involved in sensor discovery and configuration. In the future, we expect to address the open challenges discussed in Section \ref{sec:Open_Challenges}. In addition, we expect to integrate our solution with other existing solutions such as MOSDEN \cite{ZMP009} and OpenIoT \cite{P377}. The functionality provided by CADDOT can improve these solutions in a major way.

\begin{acknowledgement}
Authors acknowledge support from SSN TCP, CSIRO, Australia and ICT  Project, which is co-funded by the European Commission under seventh framework program, contract number FP7-ICT-2011-7-287305-OpenIoT. The Author(s) also acknowledge help and contributions from The Australian National University.
\end{acknowledgement}



%
%

%
%

\bibliographystyle{abbrv}
\bibliography{Bibliography}

\end{document}